\documentclass[preprint,12pt,authoryear]{elsarticle}

\usepackage{amsmath,amssymb}
\usepackage{booktabs}
\usepackage{multirow}
\usepackage{url}
\usepackage[hidelinks]{hyperref}

\newcommand{\dset}{\textsc{SatoyamaCT}}

\journal{Computers and Electronics in Agriculture}

\begin{document}

\hfuzz=3pt

\begin{frontmatter}

\title{\dset{}: A Multi-Axis Night-IR Camera-Trap Benchmark for
Monitoring Crop-Damaging Wildlife in Japanese Agroforestry}

\author[a]{Keito Inoshita\corref{cor1}}
\ead{inosita.2865@gmail.com}

\author[b]{Kohei Hisayama\fnref{fnpa}}
\fntext[fnpa]{Present address: Wakayama Fruit Tree Experiment Station,
Wakayama, Japan.}

\author[c]{Haruto Sugeno}

\author[d,e]{Kota Nojiri}

\cortext[cor1]{Corresponding author}

\affiliation[a]{organization={Faculty of Business and Commerce, Kansai University},
               city={Suita}, state={Osaka}, country={Japan}}

\affiliation[b]{organization={Graduate School of Bioresources, Mie University},
               city={Tsu}, state={Mie}, country={Japan}}

\affiliation[c]{organization={Faculty of Symbiotic Systems Science, Fukushima University},
               city={Fukushima}, state={Fukushima}, country={Japan}}

\affiliation[d]{organization={Graduate School of Agricultural and Life Sciences,
               The University of Tokyo},
               city={Bunkyo}, state={Tokyo}, country={Japan}}

\affiliation[e]{organization={The University Museum, The University of Tokyo},
               city={Bunkyo}, state={Tokyo}, country={Japan}}

\begin{abstract}
Crop and forest damage from sika deer, wild boar, and Japanese macaque is a serious economic problem in Japanese satoyama, where farmland and forest intermingle. Camera traps enable scalable monitoring, yet existing benchmarks evaluate recognition in-distribution, rarely prioritize night infrared imagery, and do not jointly address regional domain shift, novel-species detection, and uncertainty-based triage on a single dataset. We introduce the Satoyama Camera Trap Dataset (\dset{}), $12{,}642$ expert-verified crops from night-IR-dominant camera traps across three satoyama regions. An iterative annotation protocol combining BioCLIP embeddings with confidence-ordered confirmation reduced expert effort while all labels were verified by domain ecologists; species-level inter-annotator agreement reached Cohen's $\kappa = 0.900$. A multi-axis protocol jointly evaluates domain generalization across region, camera placement, and illumination; open-set novel-species detection; and selective prediction, all under capture-event-based leakage control. Difficulty separates into distinct failure modes: a data-inherent regional gap of $16$ to $27$ percentage points persists in Wakayama across four backbones and domain-generalization methods including CORAL, DANN, and GroupDRO, and open-set detection reaches AUROC $0.93$ to $0.96$ overall yet degrades jointly with classification in Wakayama. Selective prediction recovers Wakayama accuracy from $0.593$ to $0.815$ at $50\%$ coverage, supporting an uncertainty-aware triage workflow for practical pest monitoring in agroforestry.
\end{abstract}


\end{frontmatter}

\section{Introduction}
\label{sec:intro}

Satoyama, the traditional mosaic of farmland, coppiced woodland, and rural settlement that is characteristic of the Japanese countryside, is a landscape in which human activity and wildlife habitat are closely intertwined, and it is a distinctively Japanese and ecologically important setting for studying the interaction between people and ecosystems and for managing it sustainably. In Japanese satoyama, where agricultural fields and managed forests are interspersed with wildlife habitat, sika deer (\textit{Cervus nippon}), wild boar (\textit{Sus scrofa}), and Japanese macaque (\textit{Macaca fuscata}) cause serious and recurring damage to crops and timber production~\cite{tuia2022perspectives,snapshotjapan2023}. Continuous population monitoring of these pest species is essential for evidence-based wildlife management and damage prevention, yet traditional survey methods are labor-intensive and cannot cover large agroforestry landscapes at operational scale. Camera traps, or motion-triggered passive sensors, enable noninvasive and continuous recording of wildlife presence and activity across wide areas, and deep learning approaches to automated species recognition have advanced rapidly to process the hundreds of thousands to millions of records such deployments generate~\cite{norouzzadeh2018automatically,tabak2019machine}. The public release of large-scale datasets such as Snapshot Serengeti~\cite{swanson2015snapshot} has enabled automated identification at accuracy levels approaching human performance under controlled conditions~\cite{norouzzadeh2018automatically}, and automated species identification is increasingly becoming foundational technology for operational pest monitoring in agricultural and forest management~\cite{tuia2022perspectives}.

However, most reported accuracy figures are measured under conditions homogeneous with the training data, whereas in real deployments models are transferred to regions, seasons, and sensor types different from those used during training. This domain generalization gap constitutes the principal operational bottleneck. Beery et al.~\cite{beery2018recognition} demonstrated quantitatively that recognition accuracy drops substantially at new camera locations not seen during training, and WILDS~\cite{koh2021wilds} formalized inter-location distribution shift as a canonical out-of-distribution generalization challenge. Moreover, the most valuable ecological information is often concentrated among rare species or novel invasive individuals. Accordingly, operational models are expected to surface such unknown species as novel rather than forcing them into known categories, and to delegate uncertain predictions to expert review in an uncertainty-aware manner~\cite{miao2021iterative,openinsect2025}. Additionally, because many wildlife species are nocturnal, the majority of crops captured in practice are recorded as infrared monochrome. Night-infrared imaging, domain shift, and long-tailed species distributions thus form three simultaneous challenges in real deployment.

Despite these operational requirements, existing camera trap benchmarks do not address these challenges in an integrated manner. Four specific limitations are common. First, most datasets mix daytime and nighttime data without treating night infrared recording as the primary evaluation axis, and do not span both photographs and video on a single dataset~\cite{beery2021iwildcam,tabak2019machine}. Second, they lack orthogonal design along camera-placement stratification and domain-shift axes; region, modality, and camera placement remain confounded, making it impossible to attribute performance degradation to a specific axis. Third, evaluation protocols do not incorporate reliability assessment in the form of open-set novel species detection, calibration, or selective prediction, so difficulty is measured purely by mean accuracy. Fourth, geographic coverage is biased toward large protected areas in Africa, North America, and Europe, with mid-scale agricultural and forest landscapes in Asia, and in particular Japanese satoyama environments, being markedly underrepresented~\cite{snapshotjapan2023}. As a result, the axis from which operational difficulty arises remains unresolved.

In this study, we construct the Satoyama Camera Trap Dataset (\dset{}), an integrated benchmark that fills these gaps. \dset{} comprises $12{,}642$ expert-verified crops from camera traps deployed in three Japanese satoyama regions, Mie, Wakayama, and Shizuoka, spanning two modalities: photographs from Mie ($4{,}736$ crops) and video from Wakayama and Shizuoka ($7{,}906$ crops). Three characteristics define the dataset. First, the large majority ($9{,}583$ crops; approximately $76\%$) are captured in night infrared monochrome, recording the activity of nocturnal wildlife. Second, pest species such as sika deer and raccoon dog (\textit{Nyctereutes procyonoides}) are dominant, forming a strongly long-tailed distribution that extends to rare and novel individuals, with valuable information concentrated on the minority side. Third, the Wakayama video is stratified by camera placement including broadleaf-upper, broadleaf-middle, broadleaf-lower, conifer, field, and pond, embedding realistic distribution shifts across region, modality, camera placement, and illumination. Crops are annotated through a model-in-the-loop semi-automatic annotation protocol, in which animal detection using MegaDetector~\cite{beery2019efficient} is followed by BioCLIP-embedding-based clustering and confidence triage, and all final labels are verified by domain ecology experts. Label quality is additionally assessed through independent dual annotation by two experts. On \dset{}, we define and jointly evaluate domain generalization across region, camera placement, and illumination, open-set novel species detection, and calibration with selective prediction, incorporating capture-event-based leakage control throughout. Systematic evaluation using linear probes, full fine-tuning, representative domain generalization methods, and out-of-distribution detection methods is conducted to empirically verify the axis from which difficulty arises. The main contributions of this paper are as follows.

\begin{enumerate}
\item[i)] An investigator-designed multi-axis evaluation protocol is developed on
author-collected night-IR camera-trap data from three Japanese satoyama regions, jointly
evaluating domain generalization across region, camera placement, and illumination shift, open-set
novel species detection, and selective prediction on a single dataset with
capture-event-based leakage control, so that failure modes existing benchmarks treat
separately are diagnosed together.

\item[ii)] An investigator-developed iterative annotation scheme is applied to all
$12{,}642$ crops, in which BioCLIP embeddings and confidence-ordered active confirmation
reduce expert labeling burden while all labels are verified by domain ecologists and an
independent second annotator, yielding an evaluation platform with verifiable label quality.

\item[iii)] Three operational findings guide automated pest monitoring in agroforestry:
generalization to one region is consistently the hardest across backbones and
domain-generalization methods; recall of known pest species that become placement minorities
improved substantially after label correction and is no longer an independent hardcore, with the remaining dominant challenge concentrating on the regional gap;
and open-set detection, though tractable in aggregate,
degrades jointly with classification in the hardest region, a compound vulnerability
invisible under single-axis evaluation.
\end{enumerate}

The remainder of this paper is organized as follows. Section~\ref{sec:related} reviews related work. Section~\ref{sec:methods} describes the dataset construction, semi-automatic annotation procedure, and multi-axis benchmark design. Section~\ref{sec:results} presents dataset statistics and systematic baseline evaluation results. Section~\ref{sec:discussion} discusses key findings and limitations, and Section~\ref{sec:conclusion} concludes with directions for future work.

\section{Related Work}
\label{sec:related}

\subsection{Camera Trap Datasets and Benchmarks}
Camera traps are a standard observational tool for biodiversity monitoring, and the advance of machine learning has prompted the successive release of large-scale datasets. Snapshot Serengeti~\cite{swanson2015snapshot} is a pioneering dataset in which approximately forty mammalian species in an African savanna were annotated through a citizen science platform. Norouzzadeh et al.~\cite{norouzzadeh2018automatically} used the same data to demonstrate that species classification, individual counting, and behavioral description can all be automated with deep learning. Subsequently, NACTI covering North America~\cite{tabak2019machine}, Snapshot Safari spanning the African continent~\cite{pardo2021snapshot}, and iWildCam aggregating dozens of worldwide camera sites~\cite{beery2020iwildcam,beery2021iwildcam} have been released, steadily expanding the geographic and taxonomic coverage of available resources. These datasets have primarily been organized as frameworks for measuring species classification accuracy.

Diversification has also progressed on the task side. AP-10K targets animal pose estimation~\cite{yu2021ap10k}, WildlifeReID-10k addresses individual re-\-iden\-ti\-fi\-ca\-tion~\cite{adam2024wildlifereid}, and Open-Insect benchmarks open-set recognition of novel insect species~\cite{openinsect2025}, extending the scope beyond classification. In Japan, Snapshot Japan 2023 has provided the first publicly available dataset conforming to internationally standardized protocols~\cite{snapshotjapan2023}, and benchmarks accommodating temporal protocols and multi-trap operations have also been reported~\cite{streamingbench2026}. These extensions indicate a growing shift of attention toward operational challenges that cannot be captured by a single accuracy metric.

However, these benchmarks share common limitations. Most mix daytime and nighttime imagery without placing night infrared recording at the center of evaluation, and do not jointly handle photographs and video on the same dataset. Geographically, the focus remains on large protected areas in Africa, North America, and Europe; no benchmark is dedicated to mid-scale agricultural and forest landscapes such as Japanese satoyama. Furthermore, a framework that simultaneously evaluates camera-placement stratification, open-set detection, and uncertainty estimation is absent across existing resources. Table~\ref{tab:related} summarizes the relationship between \dset{} and the major existing datasets. \dset{} differs from existing benchmarks in that it co-locates these axes on a single set of expert-verified data, enabling analysis that attributes difficulty to each axis individually.

\begin{table}[t]
\centering
\caption{Comparison with existing camera trap benchmarks ($\checkmark$: supported;
$\triangle$: partial; IR = night-IR primary axis; Vd = video; Hab = camera-placement stratification;
OS = open-set evaluation; Unc = uncertainty evaluation; DG = domain generalization axis;
HITL = model-in-the-loop annotation; bottom row = proposed).}
\label{tab:related}
\small
\resizebox{\linewidth}{!}{%
\begin{tabular}{lcccccccc}
\toprule
Dataset & IR & Vd & Hab & OS & Unc & DG & HITL & Region \\
\midrule
Snapshot Serengeti~\cite{swanson2015snapshot} & $\triangle$ & & & & & & & Africa \\
Caltech CT~\cite{beery2018recognition} & $\triangle$ & & & & & $\checkmark$ & & N.\ America \\
iWildCam~\cite{beery2020iwildcam,beery2021iwildcam} & $\triangle$ & & & & & $\checkmark$ & & World \\
WILDS~\cite{koh2021wilds} & $\triangle$ & & & & & $\checkmark$ & & World \\
NACTI~\cite{tabak2019machine} & $\triangle$ & & & & & & & N.\ America \\
WildlifeReID-10k~\cite{adam2024wildlifereid} & & & & $\triangle$ & & & & Multi \\
Open-Insect~\cite{openinsect2025} & & & & $\checkmark$ & & $\triangle$ & & Europe \\
Snapshot Japan~\cite{snapshotjapan2023} & $\triangle$ & & & & & & & Japan \\
Streaming CT~\cite{streamingbench2026} & $\triangle$ & & & & $\triangle$ & $\checkmark$ & & Multi \\
\midrule
\textbf{\dset{} (proposed)} & $\checkmark$ & $\checkmark$ & $\checkmark$ & $\checkmark$ & $\checkmark$ & $\checkmark$ & $\checkmark$ & Japan satoyama \\
\bottomrule
\end{tabular}%
}
\end{table}

\subsection{Domain Generalization, Modality, Open-Set Recognition, and Uncertainty}
Inter-location generalization, in which recognition accuracy drops substantially when the evaluation location was not seen during training, is a central challenge in camera trap machine learning. Beery et al.~\cite{beery2018recognition} quantitatively demonstrated this phenomenon using Caltech Camera Traps, and WILDS~\cite{koh2021wilds} adopted iWildCam as a canonical out-of-distribution generalization task. On the operational side, an efficient pipeline centered on a general-purpose animal detector~\cite{beery2019efficient}, a survey of opportunities and challenges for machine learning in monitoring~\cite{tuia2022perspectives}, and comparative evaluations of existing platforms~\cite{velez2023evaluation} have accumulated evidence that inter-location shift is an unavoidable practical problem.

Differences in modality, and in particular the distinction between still images and video, are also consequential. DeepWILD~\cite{deepwild2023} identifies challenges specific to video data, and methods for nighttime infrared-specific detection~\cite{wang2024yolov8night} and infrared data screening~\cite{irscreening2025} have been proposed. However, a framework capable of evaluating multiple axes of distribution shift simultaneously on a single dataset, including region, camera placement, modality, and illumination, does not yet exist. As long as photographs and video are handled on separate datasets in separate regions, modality effects and regional effects cannot be disentangled, and the origin of performance degradation remains unclear.

Furthermore, in real deployments models must detect unseen species as novel rather than misclassifying them as known ones. The maximum softmax probability (MSP)~\cite{hendrycks2017baseline} is a simple yet effective novelty score, and the observation that deep models are often miscalibrated~\cite{guo2017calibration} has raised awareness of calibration as a reliability concern. Open-Insect~\cite{openinsect2025} benchmarks open-set detection for insects, and probabilistic prediction calibration and validation have been discussed in ecological contexts as well~\cite{chivers2014validation}. The present work integrates metrics including AUROC, FPR@95TPR, expected calibration error (ECE), and the risk-coverage area under the risk-coverage curve (AURC) into a multi-axis benchmark, providing evaluation axes for open-set detection, calibration, and selective prediction on the same data as classification performance, thereby filling a gap that prior camera trap benchmarks have left unaddressed.

\subsection{Foundation Model Embeddings and Model-in-the-Loop Annotation}
Foundation models based on large-scale pretraining are becoming established as a feature extraction backbone for camera trap machine learning. Following CLIP~\cite{radford2021clip} and DINOv2~\cite{oquab2024dinov2}, the biology-specialized BioCLIP~\cite{stevens2024bioclip} learns embeddings aligned with the tree of life and enables zero-shot identification of unseen taxonomic groups. Language-guided adaptation to camera traps has also been proposed, including WildCLIP~\cite{gabeff2024wildclip} and CATALOG~\cite{santamaria2025catalog}. In this work, both BioCLIP and DINOv2 are used as reference baselines, and as described in the results, the difficulty of \dset{} is confirmed to be independent of backbone choice.

Reducing annotation burden has motivated model-in-the-loop annotation, in which a model and human annotators iterate interactively. Miao et al.~\cite{miao2021iterative} demonstrated that iterative annotation achieves high accuracy with few labels on long-tailed and dynamic data, and large cloud-based platforms such as Wildlife Insights~\cite{ahumada2020wildlife} have been put into operation. From an active learning perspective, frameworks for selecting cases to confirm based on coverage of the embedding space or prediction uncertainty are well established. The annotation approach adopted in this study stands in this lineage, designing a semi-automatic annotation scheme that combines BioCLIP-embedding-based clustering with active confirmation ordered by prediction confidence.

However, in camera trap dataset construction, the efficiency of labeling has received attention while an independent framework for verifying the quality of the resulting labels has often been overlooked. Night infrared video-derived crops in particular contain individuals whose species cannot be determined from a still frame, and inter-annotator disagreement is likely for such cases. In this study, a second annotator independently labels the same seed set used by the primary annotator for all $7{,}906$ video crops, and inter-annotator agreement (IAA) is computed from the $1{,}000$ overlapping crops, providing verifiable label quality assurance that complements annotation efficiency.

\section{Dataset Construction}
\label{sec:methods}

\begin{figure}[t]
\centering
\includegraphics[width=\linewidth]{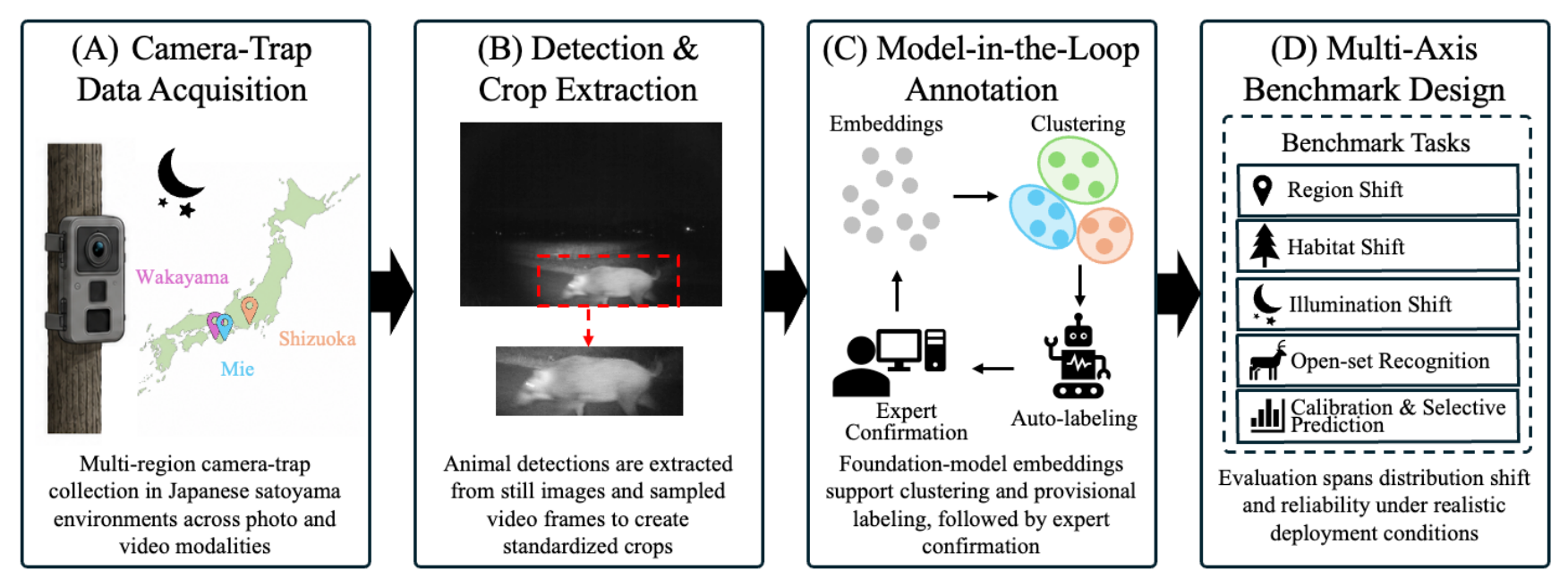}
\caption{\dset{} construction and benchmark design pipeline (A: night-IR camera traps
showing the three satoyama study regions in Japan, B: detection and crop extraction,
C: model-in-the-loop annotation, D: multi-axis benchmark); map lines delineate study
areas and do not necessarily depict accepted national boundaries.}
\label{fig:pipeline}
\end{figure}

Figure~\ref{fig:pipeline} illustrates the end-to-end construction and evaluation pipeline from data acquisition to the completed benchmark. The figure depicts four stages: night-IR camera trap capture (A), MegaDetector-based detection and crop extraction (B), model-in-the-loop annotation (C), and multi-axis benchmark design (D), and each of the following subsections corresponds to this flow.

\subsection{Data Sources and Crop Extraction}
\label{sec:sources}
\begin{figure}[t]
\centering
\includegraphics[width=0.92\linewidth]{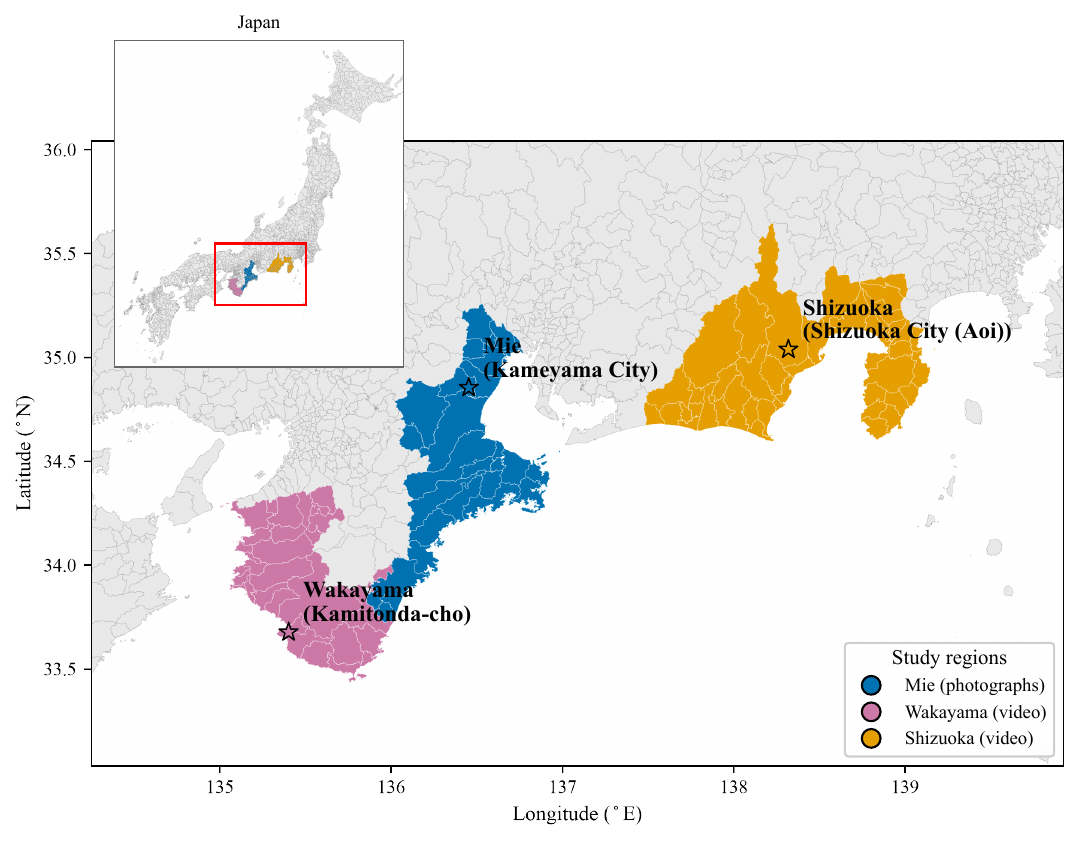}
\caption{Locations of the three satoyama study regions in central Japan: Mie
(Kameyama City; photographs), Wakayama (Kamitonda-cho; video), and Shizuoka
(Shizuoka City, Aoi Ward; video). An inset shows their position within Japan.
Stars mark the survey municipality of each region; the study did not span these
prefectures in their entirety, and precise camera-trap coordinates are withheld to
protect rare species. Map lines delineate study areas and do not necessarily depict
accepted national boundaries.}
\label{fig:sites}
\end{figure}

\begin{figure}[t]
\centering
\includegraphics[width=\linewidth]{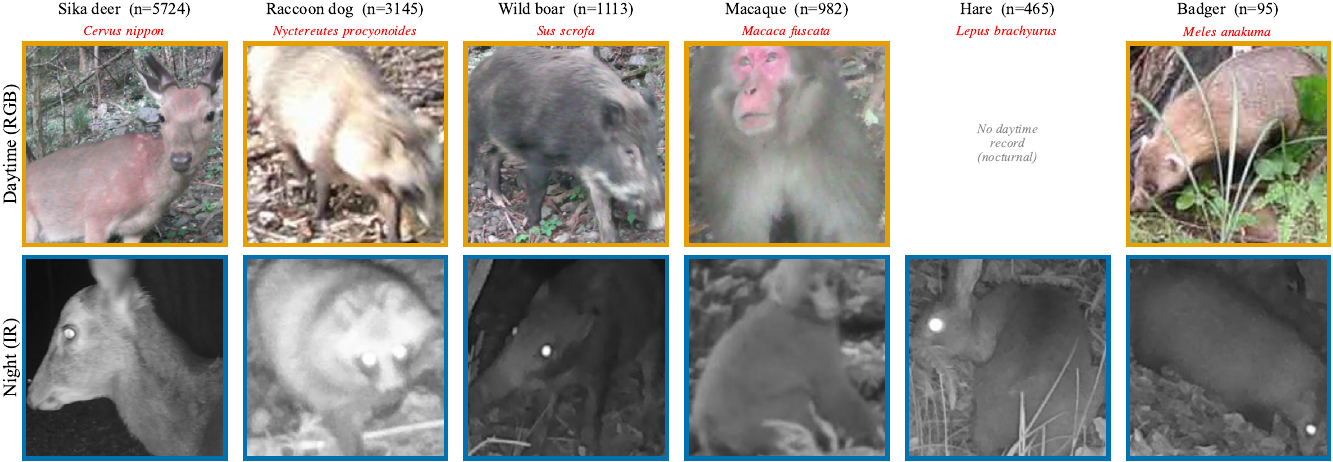}
\caption{Representative crop examples from \dset{}: each column is one species, upper rows
are daytime (RGB) and lower rows are night infrared (IR), and $n$ denotes total crop count.
The Japanese hare column has no daytime (RGB) crop because this species is
strictly nocturnal in the dataset.}
\label{fig:examples}
\end{figure}

\begin{figure}[t]
\centering
\includegraphics[width=\linewidth]{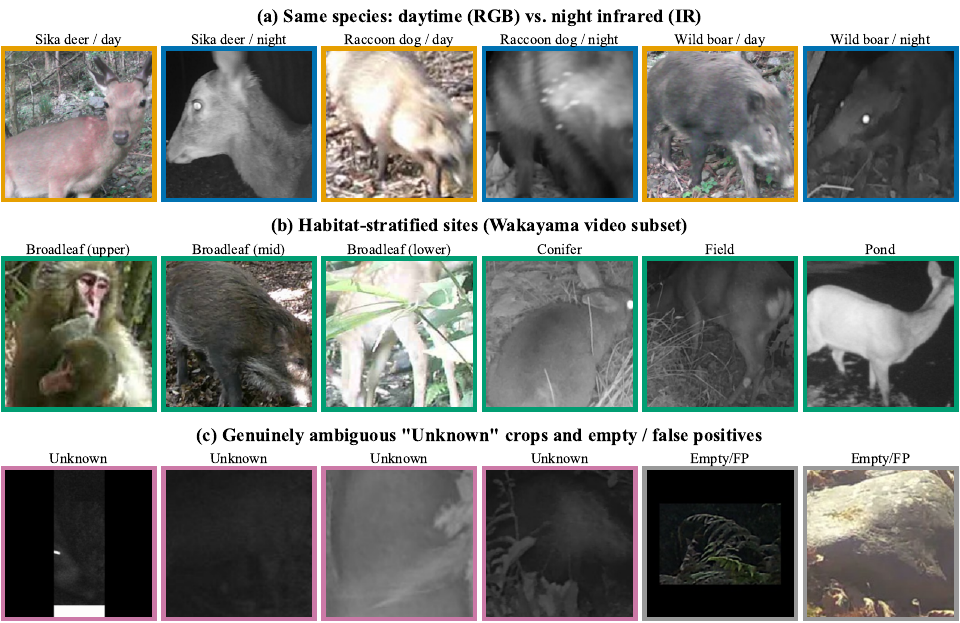}
\caption{Diversity of capture conditions (all real crops): (a) the same species under
daytime RGB versus night IR, (b) camera-placement stratification in Wakayama, (c) ambiguous unknown
instances and false detections.}
\label{fig:conditions}
\end{figure}

This subsection corresponds to panels A and B of Figure~\ref{fig:pipeline}, covering night-IR camera trap capture and detection with crop extraction. \dset{} is derived from camera traps installed in Japanese satoyama environments and contains data from three regions and two modalities. The geographic locations of the three study regions are shown in Figure~\ref{fig:sites}. The three data sources are as follows. Mie (Kameyama City) contributes photographs consisting of native still images up to $6144 \times 3456$ pixels triggered by motion detection, the majority of which are night infrared monochrome. Shizuoka (Shizuoka City) contributes MP4 video clips, and Wakayama (Kamitonda-cho) contributes AVI video clips; data were collected at the municipalities named above rather than across these prefectures in their entirety. The Wakayama data are stratified by habitat, covering broadleaf-upper, broadleaf-middle, broadleaf-lower, conifer, field, and pond. These six correspond to camera deployment points within a single region, where broadleaf-upper, broadleaf-middle, and broadleaf-lower are three positions within one broadleaf stand and conifer, field, and pond are nearby points differing in local vegetation; the ``habitat'' axis below therefore denotes this within-region variation in placement and local vegetation rather than geographically separated habitats. The photographs were confirmed, from resolution, EXIF metadata, and camera model information, to be native still images captured in photograph mode rather than frames extracted from video. The contrast between photographs and video therefore constitutes a genuine modality difference. Because the region also differs, region and modality are structurally confounded, and this confounding is addressed explicitly in Section~\ref{sec:tasks} and Section~\ref{sec:discussion}. Representative real crops are shown in Figure~\ref{fig:examples}, and the diversity of capture conditions, spanning daytime-nighttime infrared, habitat, and ambiguous cases, is shown in Figure~\ref{fig:conditions}.

MegaDetector (MDV6-yolov9-c) was applied to each image and each video frame to extract animal bounding boxes. All still images were processed in their entirety, while video clips were sparsely sampled at five equally spaced relative positions from $0.1$ to $0.9$ using ffmpeg. Animal boxes with detection confidence at or above $0.5$ were saved as crops. Infrared versus daytime color classification was determined by computing the mean absolute color difference across RGB channels: crops falling below a threshold of $8/255$ were classified as infrared monochrome. This processing yielded $4{,}736$ crops from photographs and $7{,}906$ crops from video, totaling $12{,}642$ crops.

\subsection{Model-in-the-Loop Semi-Automatic Annotation}
\label{sec:annot}
This subsection corresponds to panel C of Figure~\ref{fig:pipeline}. Manually classifying all $12{,}642$ crops was infeasible, and video-derived crops contain individuals whose species cannot be determined from still frames alone. A model-in-the-loop semi-automatic annotation scheme combining foundation model embeddings with active learning was therefore designed to reduce the burden on domain expert annotators. First, each crop was embedded into $512$ dimensions by BioCLIP, a species-recognition foundation model, and divided into fine-grained clusters using $k$-means, so that visually similar individuals were grouped adjacently within each cluster by intra-cluster similarity ordering. Second, clusters were presented as grids on a web tool that allowed experts to select same-species crops simultaneously by dragging a rectangle, to distinguish mixed cases with separate labels, and to exclude detections that did not contain the target species. Third, a stratified sample of $1{,}000$ representative crops near cluster centroids was extracted as seed crops for labeling; experts could add species not present in an initial list at any point. Fourth, a multiclass logistic regression was trained on the seed labels and the confirmed photograph labels, and prediction labels with confidence scores were assigned to all video crops. Finally, a confirmation site presenting crops sorted by ascending prediction confidence per predicted species was shown to experts, who confirmed high-confidence predictions in bulk and focused corrections on low-confidence and ambiguous cases.

Automatic predictions achieved a hit rate of $54.5\%$ on the $6{,}906$ crops excluding seeds, demonstrating a substantial reduction in manual review workload through model assistance. All final labels were confirmed by domain experts. Because confirmation proceeds in order of model prediction confidence, there is a concern that high-confidence but incorrect samples may be overlooked due to circular bias. Expert correction rates were therefore measured by confidence band, using the $6{,}906$ non-seed crops as the denominator. In the low-confidence band $[0, 0.5)$, $68.0\%$ of non-seed crops were corrected, whereas in the high-confidence band $[0.9, 1)$ the correction rate was only $1.15\%$ with a $95\%$ confidence interval of $[0.53, 2.48]\%$, confirming that high-confidence automatic labels are largely correct and that residual circular bias is small. False detections with no animal present were incorporated as an explicit classification class, so that detector errors were automatically routed and excluded through the labeling and confirmation pipeline.

\subsection{Label Quality Verification: Inter-Annotator Agreement}
\label{sec:iaa}
To assure label quality independently of annotation efficiency, a second annotator independently labeled the same seed set of $1{,}000$ crops used by the primary annotator responsible for gold labels across all $7{,}906$ video crops. IAA was computed from the $1{,}000$ overlapping crops. In a full evaluation covering all $13$ categories, raw agreement was $77.7\%$ and Cohen's $\kappa = 0.687$, corresponding to substantial agreement under the Landis-Koch criterion. Restricting the evaluation to species only, excluding unknown and false detection categories, yielded $n = 639$ pairs with raw agreement $94.8\%$ and $\kappa = 0.900$, reaching almost perfect agreement, indicating that when species identification was possible the labels are highly consistent.

\begin{figure}[t!]
\centering
\includegraphics[width=\linewidth]{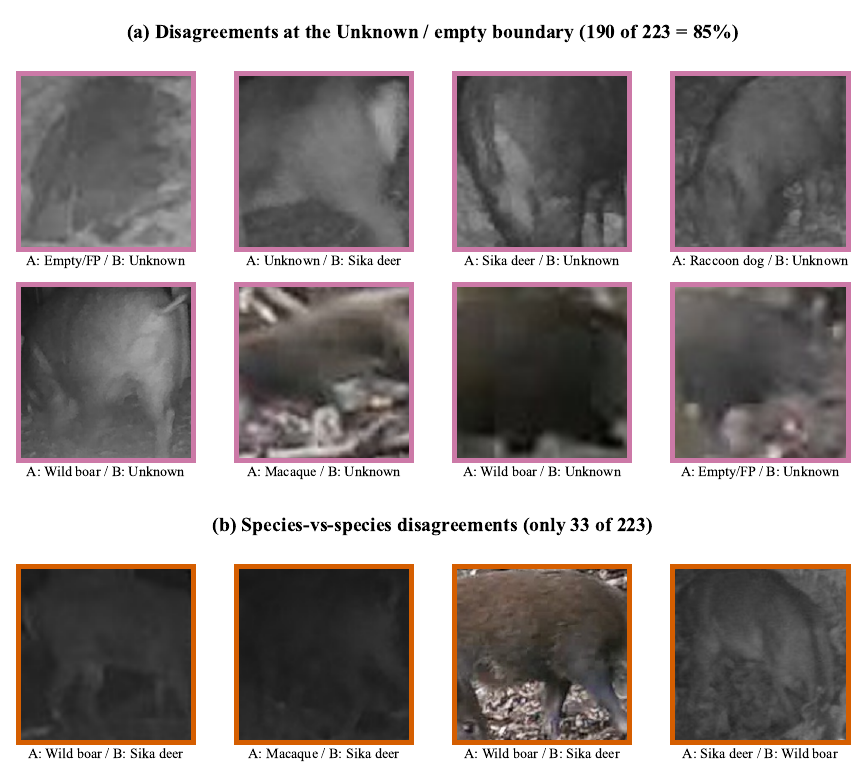}
\caption{Examples of disagreements in independent dual annotation: (a) $85\%$ of
disagreements involve the unknown or false-detection boundary, (b) pure inter-species
confusion accounts for only $33$ cases. A and B denote the two independent expert
annotators.}
\label{fig:iaadisagree}
\end{figure}

Figure~\ref{fig:iaadisagree} presents real crop examples of disagreements from the independent dual annotation. The gap between overall $\kappa$ and species-only $\kappa$ stems from the handling of difficult cases. Of $223$ disagreements, $85.2\%$ ($190$ cases) involved unknown or false detection categories, while pure inter-species confusion accounted for only $33$ cases. The most frequent disagreement pattern was primary annotator labeling unknown and secondary annotator labeling sika deer ($50$ cases), a characteristic asymmetry arising when one annotator withholds judgment and the other makes an active species identification for a crop with an indistinct subject in night IR. Stratified agreement was Wakayama $\kappa = 0.690$ and Shizuoka $\kappa = 0.666$; for illumination, IR-positive crops yielded $\kappa = 0.645$ compared with $\kappa = 0.720$ for IR-negative crops. The lowest agreement was observed for the Wakayama field habitat at $\kappa = 0.396$.

The foregoing results demonstrate that species-level labels achieve almost perfect agreement ($\kappa = 0.900$), supporting the reliability of the gold labels. At the same time, the difficulty of assigning unknown and false detection labels, which is what depresses the overall $\kappa$, is itself an intrinsic ambiguity of the night-IR and minority-species context, and the design choice to retain these cases as targets for open-set and selective prediction evaluation is consistent with this observation. IAA thereby independently corroborates that the difficulty structure described subsequently derives from the inherent properties of the data rather than from arbitrariness in labeling.

\subsection{Species Taxonomy and Leakage Control}
\label{sec:taxonomy}
This subsection and the following one correspond to panel D of Figure~\ref{fig:pipeline}, covering multi-axis benchmark design. Confirmed labels comprise $19$ classes including mammals, birds, an other category, unknown, and false detection. The species distribution is strongly skewed: sika deer and raccoon dog together account for approximately $70\%$ of all crops, while many species number in the tens of examples or fewer, forming a long tail. In the benchmark, labels are organized into the following roles: the five species with sufficient sample counts, namely raccoon dog, sika deer, wild boar, Japanese macaque, and Japanese hare (\textit{Lepus brachyurus}), are designated known; remaining real species present in small numbers are designated novel; unknown cases are designated uncertain; and false detections without animals are designated junk.

\label{sec:leak}
Because camera traps take bursts of near-duplicate frames upon each trigger event, a random crop-level split would leak near-duplicates between the training and evaluation sets. To prevent this, each crop was assigned a unified timestamp derived from its filename or EXIF metadata, and within each site, gaps of at most $120$ seconds were aggregated into a burst as the leakage unit, and gaps of at most $30$ minutes into an independent event as the counting unit. The $120$-second threshold conservatively covers the typical burst and re-trigger interval of $5$ to $60$ seconds for camera traps; for video data, one clip is treated as one group. The primary splits crossing region and camera-placement boundaries, specifically region leave-one-site-out (LOSO) and placement LOSO, are structurally free of leakage because they involve different cameras and time periods. The open-set evaluation, which splits training and evaluation within the same region, applies a clip-level held-out partition for known classes; the illumination shift split excludes ambiguous clips where color classification may reverse around twilight, removing group-level leakage.

\subsection{Benchmark Tasks and Evaluation Metrics}
\label{sec:tasks}
Reflecting the operational question of whether a model deployed to a new location, sensor, or environment will function in practice, multiple domain shift axes and reliability evaluation are defined as follows. In region domain generalization, or region-LOSO, one of Mie, Wakayama, and Shizuoka is held out for evaluation and the remaining two are used for training. Because region and modality are confounded, this setting models real deployment to a new region with a new sensor. In camera-placement domain generalization, or placement-LOSO, one deployment point within the Wakayama video is held out for evaluation, measuring a pure placement shift within the same region and modality. In open-set recognition, models trained on known species are evaluated on their ability to detect novel rare species as new; an IR-only variant is also reported to eliminate the illumination confound. The illumination shift task isolates a pure illumination domain gap by contrasting daytime color and night infrared for sika deer alone, avoiding species confounding. Evaluation metrics include accuracy and macro-F1 for species recognition, ECE for calibration, AUROC and FPR@95TPR for open-set detection, and AURC for selective prediction. Note that the placement code ceymor corresponds to an internal location code for a specific installation site in Wakayama that does not reduce to a standard habitat category; because it is a small fold with $n = 75$, results for this split are marked with $\dagger$ throughout.

\section{Dataset Statistics and Benchmark Results}
\label{sec:results}

\subsection{Dataset Composition}
\label{sec:stats}

\begin{figure}[t]
\centering
\includegraphics[width=0.72\linewidth]{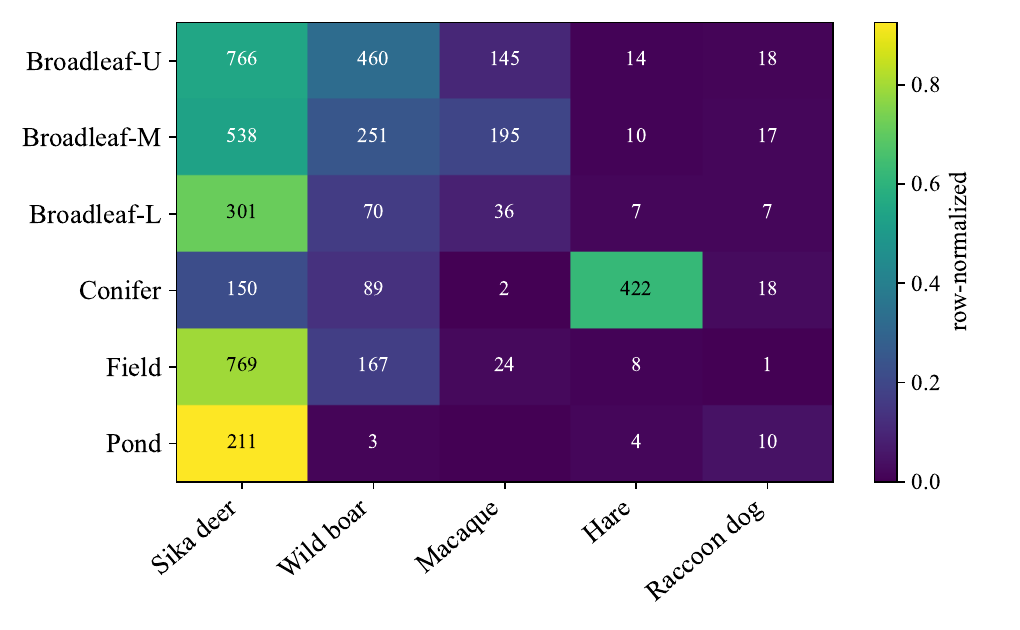}
\caption{Species composition by deployment point in Wakayama (video), showing clear community
variation across placements that makes camera placement a natural domain-shift axis.}
\label{fig:habitat}
\end{figure}

\begin{figure}[t]
\centering
\includegraphics[width=\linewidth]{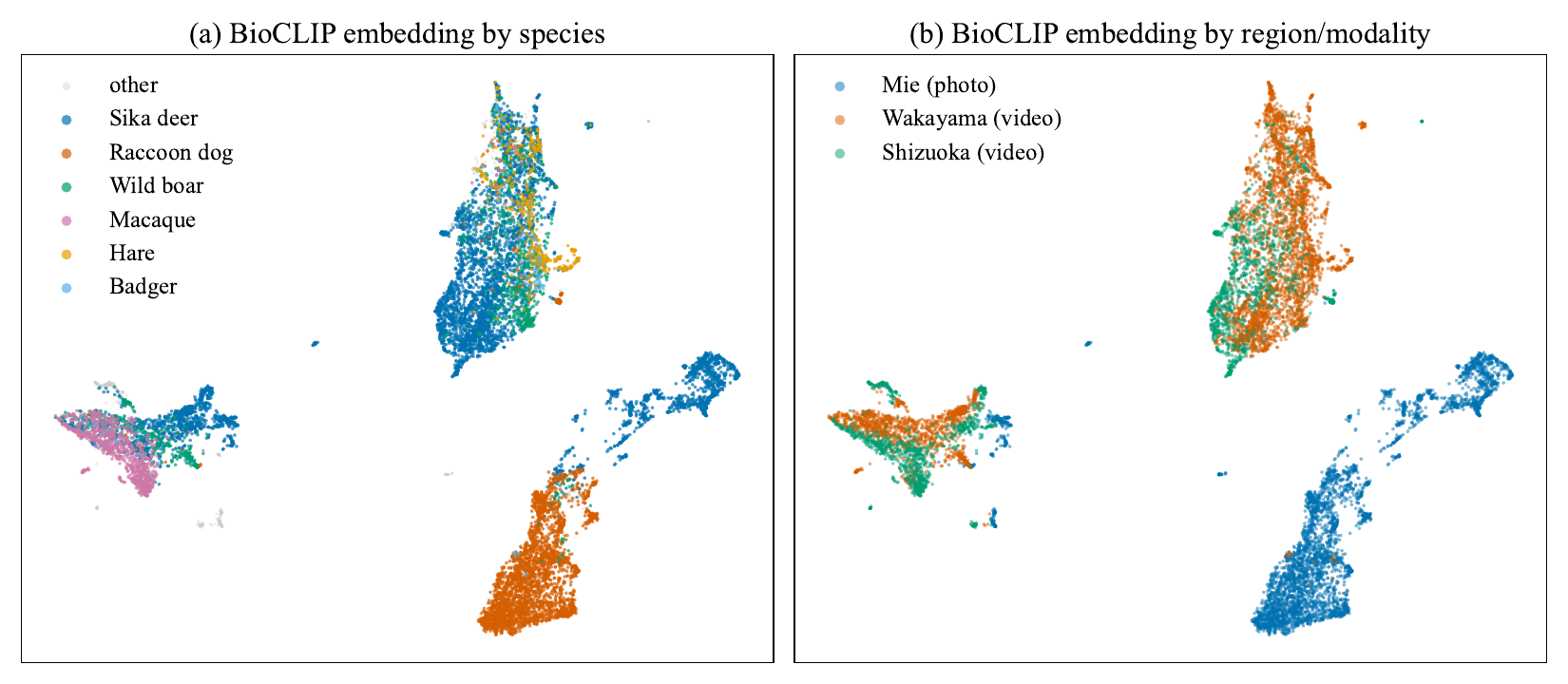}
\caption{UMAP of BioCLIP embeddings: (a) by species, (b) by region and modality, showing
the region-modality structural distribution difference and the overlap of rare-species
clusters.}
\label{fig:umap}
\end{figure}

Table~\ref{tab:compose} summarizes the dataset composition. \dset{} comprises $12{,}642$ expert-verified crops covering photographs ($4{,}736$) and video ($7{,}906$) from three regions: Wakayama ($5{,}454$), Mie ($4{,}736$), and Shizuoka ($2{,}452$). In terms of capture conditions, $9{,}583$ crops ($75.8\%$) were captured in night infrared monochrome and $3{,}059$ ($24.2\%$) in daytime color; the dominance of night infrared recording is a primary characteristic of this dataset. Photographs originate from the Mie still images and video from Wakayama and Shizuoka, so modality and region are structurally confounded; the goal of multi-axis evaluation is to decompose this confounding axis by axis. As illustrated in Figure~\ref{fig:habitat}, the Wakayama video is stratified by camera placement, and community composition varies clearly across habitats, making camera placement a natural domain-shift axis within the same region and modality. The UMAP visualization of BioCLIP embeddings shown in Figure~\ref{fig:umap} reveals the structural regional-modality distribution difference and the overlap of rare-species clusters, indicating the structural origin of the benchmark's difficulty.

\begin{table}[t]
\centering
\caption{Dataset composition: crop counts by region and modality.}
\label{tab:compose}
\small
\begin{tabular}{lllr}
\toprule
Modality & Region & Format & Crops \\
\midrule
Photo & Mie (Kameyama City) & 21\,MP still images & $4{,}736$ \\
Video & Wakayama & AVI (placement stratified) & $5{,}454$ \\
Video & Shizuoka & MP4 & $2{,}452$ \\
\midrule
\multicolumn{3}{l}{Night IR / Daytime color} & $9{,}583$ / $3{,}059$ \\
\multicolumn{3}{l}{Total} & $\mathbf{12{,}642}$ \\
\bottomrule
\end{tabular}
\end{table}

As Table~\ref{tab:species} shows, the species distribution is strongly long-tailed. Sika deer ($5{,}724$) and raccoon dog ($3{,}145$) are dominant, together accounting for approximately $70\%$ of all crops, followed by wild boar ($1{,}113$) and Japanese macaque ($982$), while unknown crops number $475$ and many species number in the tens of examples or fewer. Assigning the five species with sufficient sample counts as known, real species with tens of examples or fewer as novel, unknown as uncertain, and false detections as junk makes it possible to simultaneously support classification, open-set, and uncertainty evaluation on the same dataset. A key characteristic of the benchmark is that the class prior distribution differs substantially across regions: Mie is dominated by raccoon dog, Wakayama by sika deer and wild boar, and Shizuoka by sika deer and Japanese macaque, making region a natural testbed for label-shift evaluation.

\begin{table}[t]
\centering
\caption{Species distribution across all $12{,}642$ crops and $19$ classes. Role:
K = known, N = novel, U = uncertain, J = junk.}
\label{tab:species}
\small
\begin{tabular}{lrl}
\toprule
Species & Crops & Role \\
\midrule
Sika deer (\textit{Cervus nippon}) & $5{,}724$ & K \\
Raccoon dog (\textit{Nyctereutes procyonoides}) & $3{,}145$ & K \\
Unknown & $475$ & U \\
Wild boar (\textit{Sus scrofa}) & $1{,}113$ & K \\
Japanese macaque (\textit{Macaca fuscata}) & $982$ & K \\
False detection (no animal) & $349$ & J \\
Japanese hare (\textit{Lepus brachyurus}) & $465$ & K \\
Japanese badger (\textit{Meles anakuma}) & $95$ & N \\
Oriental turtle dove (\textit{Streptopelia orientalis}) & $57$ & N \\
Chinese bamboo partridge (\textit{Bambusicola thoracicus}) & $37$ & N \\
White's thrush (\textit{Zoothera aurea}) & $33$ & N \\
Raccoon (\textit{Procyon lotor}) & $43$ & N \\
Red fox (\textit{Vulpes vulpes}) / Japanese marten (\textit{Martes melampus}) & $17$ / $18$ & N \\
Birds & $40$ & N \\
Masked palm civet (\textit{Paguma larvata}) & $21$ & N \\
Other / Cat (\textit{Felis silvestris catus}) / Varied tit (\textit{Sittiparus varius}) & $21$ / $5$ / $2$ & N/N \\
\bottomrule
\end{tabular}
\end{table}

\subsection{Reference Baselines and the Regional Gap}
\label{sec:baselines}

Two reference baselines of contrasting character are provided on the BioCLIP embeddings. The nearest-class prototype baseline P requires no training and judges solely from the geometry of the embedding space, serving as a weak reference. The linear probe L applies a multiclass logistic regression as a lightweight trained head and is used as the primary indicator. The difference between the two reveals how much of the difficulty can be recovered by adding a light decision boundary to a fixed representation. Evaluation metrics are accuracy, macro-F1, and ECE, and $95\%$ bootstrap confidence intervals on the test set are reported alongside point estimates. Table~\ref{tab:dg} presents results across three split families.

For the region-LOSO family, which confounds region and modality, linear probe accuracy is $0.839$ for Mie and $0.816$ for Shizuoka, whereas Wakayama reaches only $0.593$, a gap of approximately $25$ percentage points below Mie. Even with the prototype baseline, Wakayama ($0.492$) falls below both Mie ($0.733$) and Shizuoka ($0.610$), indicating that the regional gap already appears at the stage of embedding geometry, before any decision boundary is learned. The non-overlapping confidence intervals across regions confirm that this difference cannot be explained by random variation. Because this compound axis contains both a modality difference and a regional difference, a video-only region-LOSO that fixes modality is also reported. The gap persists: Wakayama $0.531$ versus Shizuoka $0.836$, demonstrating that the performance degradation is not attributable to the photograph-versus-video modality difference alone. Further fixing both region and modality, the placement-LOSO within Wakayama shows that accuracy varies substantially across placements, from $0.661$ for broadleaf-upper to $0.472$ for field, confirming that camera placement constitutes a natural distribution-shift axis even within the same region and modality. From a calibration perspective, the Wakayama linear probe ECE of $0.078$ exceeds Mie's $0.054$, indicating that the validity of confidence is already degraded in the hardest region at this stage, not only accuracy. Note that macro-F1 degenerates in folds containing few classes, so accuracy, which is independent of the number of present species, is used as the primary difficulty indicator for inter-regional comparisons. The regional gap is thus consistently observed across the three orthogonal split families of compound, modality-fixed, and region- and modality-fixed (the placement-LOSO within Wakayama, in which region and modality are held constant while the deployment point varies). The following subsections verify whether this gap is attributable to the weakness of a specific model or optimization method, or whether it is a data-inherent property.

\begin{table}[t]
\centering
\caption{Domain generalization baselines (known 5 species). L = linear probe,
P = prototype. Numbers in brackets are $95\%$ bootstrap confidence intervals.
$\dagger$ = macro-F1 is unreliable for folds with few classes.}
\label{tab:dg}
\small
\begin{tabular}{lrrrl}
\toprule
Split & P-acc & L-acc [95\% CI] & L-mF1 [95\% CI] & L-ECE \\
\midrule
\multicolumn{5}{l}{\emph{Region LOSO (region $\otimes$ modality)}} \\
\quad Mie (photo) & 0.733 & 0.839 [0.83, 0.85] & 0.517 [0.37, 0.58] & 0.054 \\
\quad Wakayama (video) & 0.492 & 0.593 [0.58, 0.61] & 0.371 [0.36, 0.39] & 0.078 \\
\quad Shizuoka (video) & 0.610 & 0.816 [0.80, 0.83] & 0.491 [0.47, 0.51] & 0.065 \\
\midrule
\multicolumn{5}{l}{\emph{Video-only region LOSO (modality fixed)}} \\
\quad Wakayama & $-$ & 0.531 & 0.405 & 0.073 \\
\quad Shizuoka & $-$ & 0.836 & 0.539 & 0.112 \\
\midrule
\multicolumn{5}{l}{\emph{Placement LOSO (Wakayama)}} \\
\quad Broadleaf-upper & 0.552 & 0.661 [0.64, 0.69] & 0.499 [0.46, 0.53] & 0.055 \\
\quad Broadleaf-middle & 0.505 & 0.633 [0.60, 0.66] & 0.456 [0.42, 0.49] & 0.039 \\
\quad Broadleaf-lower & 0.499 & 0.614 [0.57, 0.66] & 0.478 [0.42, 0.54] & 0.078 \\
\quad Conifer & 0.421 & 0.575 [0.54, 0.61] & 0.395 [0.35, 0.44] & 0.103 \\
\quad Field & 0.409 & 0.472 [0.44, 0.50] & 0.341 [0.30, 0.38] & 0.082 \\
\quad Pond$\dagger$ & 0.557 & 0.732 [0.68, 0.79] & 0.328 [0.26, 0.40] & 0.121 \\
\quad ceymor$\dagger$ & $-$ & 0.760 [0.67, 0.85] & 0.489 [0.29, 0.58] & 0.153 \\
\bottomrule
\end{tabular}
\end{table}

\subsection{The Regional Gap Is Independent of Backbone Choice and Generalization Method}
\label{sec:backbone}

The regional gap is verified sequentially across three axes: backbone, full fine-tuning, and domain generalization method. Regarding backbone, Table~\ref{tab:backbone} presents region-LOSO results for four configurations whose pretraining properties differ substantially: a frozen BioCLIP probe with biology-specialized contrastive language pretraining, a DINOv2 probe with powerful self-supervised representations, a full fine-tuning of an ImageNet-pretrained ViT-B/16, and a full fine-tuning of the BioCLIP encoder itself. Across all four configurations, which vary in specialization and training paradigm, Wakayama is consistently the hardest region by $16$ to $27$ percentage points, and the gap between Wakayama and the mean of Mie and Shizuoka falls in the range $\Delta_W = -0.162$ to $-0.270$. Specifically, DINOv2, which achieves the highest within-domain accuracy among the four configurations, still yields only $0.803$ for Wakayama, and the BioCLIP full fine-tuning reaches $0.628$. Improving representation quality therefore leaves most of the Wakayama gap intact, confirming that the difficulty is a data-inherent regional domain shift independent of backbone choice.

\begin{figure}[t!]
\centering
\includegraphics[width=0.80\linewidth]{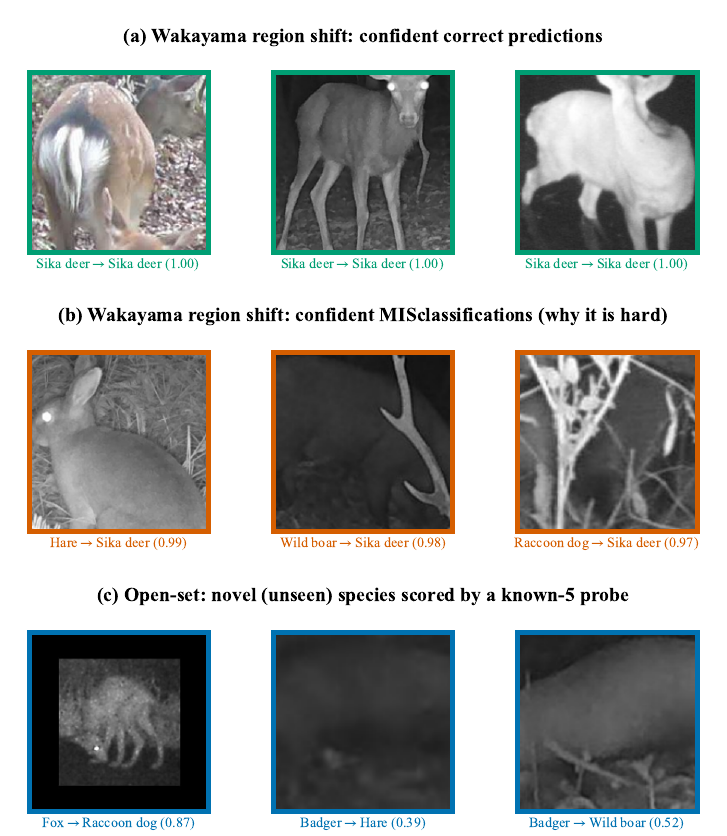}
\caption{Qualitative examples of regional domain shift with Wakayama as the evaluation
split (frozen BioCLIP embeddings plus logistic probe; all real crops). (a) High-confidence
correct predictions. (b) High-confidence failures: under Wakayama shift, several
minority species (Japanese hare, wild boar, raccoon dog) are confidently absorbed into
the dominant sika deer class at $p \approx 0.97$--$0.99$. (c) Open-set examples: novel
species unseen during training are pulled toward one of the five known classes.}
\label{fig:qualshift}
\end{figure}

Figure~\ref{fig:qualshift} presents qualitative real-crop examples of how this regional gap manifests when Wakayama is the evaluation region. The model produces some correct high-confidence predictions, but also confidently absorbs several minority species (Japanese hare, wild boar, raccoon dog) into the dominant sika deer class at $p \approx 0.97$--$0.99$ while maintaining high confidence. From an open-set perspective, novel species unseen during training are strongly attracted toward one of the five known classes and misread. These failures are not characterized by uniformly diffuse predictions but by high-confidence absorption into the wrong class, directly linking to the calibration and selective prediction challenges described in Section~\ref{sec:reliability}.

\begin{table}[t]
\centering
\caption{Accuracy and macro-F1 under region LOSO for four backbone configurations.
$\Delta_W$: accuracy gap between Wakayama and the mean of Mie and Shizuoka.
The boldface highlights the best results.}
\label{tab:backbone}
\small
\resizebox{\linewidth}{!}{%
\begin{tabular}{lrrrrrr}
\toprule
Backbone & Mie & Mie & Wakayama & Wakayama & Shizuoka & $\Delta_W$ \\
 & acc & mF1 & acc & mF1 & acc & (acc) \\
\midrule
BioCLIP-probe & 0.839 & 0.517 & 0.593 & 0.371 & 0.816 & $-0.235$ \\
DINOv2-probe  & \textbf{0.975} & \textbf{0.682} & \textbf{0.803} & \textbf{0.616} & \textbf{0.955} & $-0.162$ \\
ImageNet-ViT (FT) & 0.884 & 0.551 & 0.725 & 0.508 & 0.957 & $-0.196$ \\
BioCLIP-FT & 0.861 & 0.583 & 0.628 & 0.413 & 0.934 & $-0.270$ \\
\bottomrule
\end{tabular}%
}
\end{table}

Regarding optimization and generalization methods, the analysis next confirms that the difficulty does not stem from using only naive empirical risk minimization (ERM). Table~\ref{tab:dgmethods} reports macro-F1 for CORAL, DANN, and GroupDRO trained with head-only learning on frozen BioCLIP embeddings. Across all four methods, Wakayama macro-F1 is lowest, and the gap between Wakayama and the mean of the other regions is $0.143$ for ERM, $0.140$ for CORAL, and $0.102$ for DANN. GroupDRO, which explicitly optimizes the worst-group objective, narrows the gap from $0.143$ to $0.085$, a reduction of approximately $40\%$; yet Wakayama remains the hardest region. This conclusion is robust across random seeds: across five seeds, Wakayama is the hardest region under every seed, and the GroupDRO gap of $0.069$ is statistically significant under Welch's $t$-test ($p < 0.001$; Cohen's $d = 10.9$).

To confirm that the foregoing is not specific to head-only learning on frozen embeddings, Table~\ref{tab:e2e} presents region-LOSO accuracy for ERM, GroupDRO, and SWAD~\cite{cha2021swad} applied as full end-to-end fine-tuning of ViT-B/16. In terms of accuracy, Wakayama is robustly the hardest region across all methods: the per-seed maximum for Wakayama is at or below $0.740$, whereas the per-seed minimum for Mie and Shizuoka is at or above $0.858$, so the two ranges do not overlap. GroupDRO does not outperform ERM on Wakayama; SWAD exceeds ERM by $0.4$ percentage points ($0.728$ vs $0.724$), a difference within the cross-seed standard deviation of approximately $0.006$ and thus not statistically meaningful. Note that macro-F1 is not used for inter-regional comparison because the number of species present at test time differs across regions; accuracy, which is independent of the number of present species, is used as the difficulty indicator. The consistency with which the Wakayama gap remains even after exhausting four mutually independent interventions, backbone specialization, representation quality, distribution-robust optimization, and weight averaging, strongly suggests that the difficulty is not a surface-level optimization problem but is rooted in the data-generating distribution itself.

\begin{table}[t]
\centering
\caption{Macro-F1 under region LOSO for representative domain generalization methods
(known 5 species, head-only, seed 42). $\Delta_W$ denotes the gap between Wakayama
and the mean of the other regions. $\dagger$ = variation within $\pm 0.004$ across
independent runs. The boldface highlights the best results.}
\label{tab:dgmethods}
\small
\begin{tabular}{lrrrr}
\toprule
Method & Mie & \textbf{Wakayama} & Shizuoka & $\Delta_W$ \\
\midrule
ERM (linear) & 0.536 & \textbf{0.371} & 0.491 & $0.143$ \\
CORAL & 0.504 & \textbf{0.340} & 0.456 & $0.140$ \\
DANN$^{\dagger}$ & 0.494 & \textbf{0.359} & 0.427 & $0.102$ \\
GroupDRO$^{\dagger}$ & 0.430 & \textbf{0.375} & 0.490 & $\mathbf{0.085}$ \\
\bottomrule
\end{tabular}
\end{table}

\begin{table}[t]
\centering
\caption{Region-LOSO accuracy under ViT-B/16 full end-to-end fine-tuning (19 classes)
for ERM, GroupDRO, and SWAD. Values are mean $\pm$ standard deviation across three seeds
with unified settings. Wakayama is the hardest region under all methods.}
\label{tab:e2e}
\small
\begin{tabular}{lccc}
\toprule
Method & Mie (acc) & Wakayama (acc) & Shizuoka (acc) \\
\midrule
ERM & $0.869 \pm 0.005$ & \textbf{$0.724 \pm 0.004$} & $0.956 \pm 0.002$ \\
GroupDRO & $0.878 \pm 0.011$ & \textbf{$0.710 \pm 0.007$} & $0.953 \pm 0.004$ \\
SWAD & $0.880 \pm 0.006$ & \textbf{$0.728 \pm 0.006$} & $0.952 \pm 0.003$ \\
\bottomrule
\end{tabular}
\end{table}

\subsection{Placement-Level Recall: Improvement after Label Correction}
\label{sec:minority}

Following label correction, the picture in placement-LOSO changes substantially. In placement-LOSO, full fine-tuning improves accuracy across Wakayama placements, and macro-F1 rises to $0.53$--$0.88$ as shown in Table~\ref{tab:ft}, up from $0.38$--$0.79$ before correction. This improvement reflects the clarification of the learning task: many crops previously labeled as unknown have been assigned to specific species, enlarging the effective training signal for minority classes. That the improvement is genuine and not merely an artifact of relabeling is confirmed by the broadleaf-lower fold: this fold now contains $n = 421$ crops with all five known species present, and fine-tuned macro-F1 reaches $0.672$ averaged over three seeds, up from $0.572 \pm 0.062$ before label correction. Recall of known species that have become minority classes within a given placement therefore improves substantially, and the second source of difficulty identified before correction is correspondingly attenuated. The remaining dominant challenge is the cross-region Wakayama gap, which persists across all four backbone configurations and all optimization methods.

\begin{table}[t]
\centering
\caption{Comparison of linear probe (L) and ViT-B/16 full fine-tuning (FT) for video
splits (known 5 species). FT values are single-seed; the three key folds (Wakayama
region, broadleaf-lower, and pond) were replicated across three seeds.
$\dagger$ = small-class fold; values are for reference only.}
\label{tab:ft}
\begin{tabular}{lrrrr}
\toprule
Split & L-acc & FT-acc & L-mF1 & FT-mF1 \\
\midrule
Region: Mie (photo) & 0.839 & 0.884 & 0.517 & 0.551 \\
Region: Wakayama (video) & 0.593 & 0.725 & 0.371 & 0.508 \\
Region: Shizuoka (video) & 0.816 & 0.957 & 0.491 & 0.617 \\
\midrule
Video region: Wakayama & 0.531 & 0.722 & 0.405 & 0.491 \\
Video region: Shizuoka & 0.836 & 0.935 & 0.539 & 0.567 \\
Placement: broadleaf-upper & 0.661 & 0.866 & 0.499 & 0.803 \\
Placement: broadleaf-middle & 0.633 & 0.927 & 0.456 & 0.876 \\
Placement: broadleaf-lower & 0.614 & 0.860 & 0.478 & 0.672 \\
Placement: conifer & 0.575 & 0.784 & 0.395 & 0.606 \\
Placement: field & 0.472 & 0.907 & 0.341 & 0.764 \\
Placement: pond$\dagger$ & 0.732 & 0.956 & 0.328 & 0.563 \\
Placement: ceymor$\dagger$ & 0.760 & 0.975 & 0.489 & 0.532 \\
\bottomrule
\end{tabular}
\end{table}

\subsection{Open-Set Novel Species Detection and Illumination Shift}
\label{sec:openset}

In contrast to the classification difficulty, open-set novel species detection is a tractable axis in aggregate. The evaluation trains on the five known species and evaluates held-out known versus novel crops partitioned at the clip level. As Table~\ref{tab:openset} shows, performance depends strongly on the choice of OOD score. The prototype distance without learning achieves AUROC $0.69$, dropping to $0.59$ in the IR-only setting, because restricting novelty detection to taxonomic distance in the embedding space makes it harder to distinguish related species when color information is absent in night infrared. Learned scores that exploit the geometry of the decision boundary substantially outperform this baseline. The linear probe MSP achieves AUROC $0.825$ and FPR@95TPR $0.412$, the Energy score on DINOv2 embeddings reaches $0.934$, and the combination of full ViT-B fine-tuning with MC-Dropout entropy achieves overall AUROC $0.955$ and IR-only AUROC $0.946$, with FPR@95TPR as low as $0.149$. With an appropriate uncertainty scorer, aggregate open-set novel species detection is thus a largely tractable axis, asymmetrically easier than the regional classification gap.

This conclusion is not a product of taxonomic distance between novel species. Re-evaluating with mammals only, excluding the phylogenetically distinct bird novel species, the MSP score yields $0.832$ and the fine-tuned entropy score $0.946$, both essentially unchanged, whereas only the prototype distance collapses to $0.546$. This contrast demonstrates that the advantage of learned scorers derives from the model's own uncertainty rather than from taxonomic distance between species.

Isolating the illumination shift by fixing the species to sika deer alone, training on daytime color and evaluating on night IR yields recall $0.63$, while training on night IR and evaluating on daytime color yields $0.75$. Neither direction is perfect, and the greater difficulty of the daytime-to-nighttime direction compared with the reverse quantifies an asymmetric illumination domain gap, in which transferring to infrared imagery from color images is harder than the reverse direction.

\begin{table}[t]
\centering
\caption{Open-set evaluation (held-out known/novel discrimination). Higher AUROC and
lower FPR@95TPR are better. The boldface highlights the best results.}
\label{tab:openset}
\small
\resizebox{\linewidth}{!}{%
\begin{tabular}{llrr}
\toprule
OOD score & Setting & AUROC & FPR@95TPR \\
\midrule
Prototype distance & Overall & 0.690 & 0.773 \\
Prototype distance & IR-only & 0.590 & 0.758 \\
MSP (linear probe) & Overall & 0.825 & 0.412 \\
MSP (linear probe) & IR-only & 0.818 & 0.425 \\
Energy (BioCLIP) & Overall & 0.849 & 0.434 \\
Mahalanobis (BioCLIP) & Overall & 0.741 & 0.719 \\
Energy (DINOv2) & Overall & 0.934 & 0.198 \\
Mahalanobis (DINOv2) & Overall & 0.888 & 0.429 \\
FT + MC-Dropout entropy & Overall & \textbf{0.955} & \textbf{0.149} \\
FT + MC-Dropout entropy & IR-only & \textbf{0.946} & \textbf{0.174} \\
\bottomrule
\end{tabular}%
}
\end{table}

\subsection{Calibration, Selective Prediction, and Region-Conditional Vulnerability}
\label{sec:reliability}

Three reliability axes are evaluated in turn. First, calibration: applying temperature scaling~\cite{guo2017calibration}, the most basic form of post-hoc calibration, at each region-LOSO fold shows that ECE improves for more in-domain regions but worsens in the hardest region, Wakayama. Specifically, Mie improves from $0.047$ to $0.021$ and Shizuoka from $0.069$ to $0.031$, whereas Wakayama degrades from $0.088$ to $0.147$. Global single-temperature calibration therefore does not restore reliability under domain shift; local and distribution-aware calibration methods are an open research challenge.

Second, selective prediction: Table~\ref{tab:aurc} reports AURC for region-LOSO with the linear probe. AURC for Mie is $0.036$ and for Shizuoka $0.041$, whereas for Wakayama it is $0.192$, about five times larger. Conversely, the greatest scope for deferral is precisely where difficulty is highest: restricting to the top $50\%$ of predictions by confidence in Wakayama recovers accuracy from $0.593$ to $0.815$. Uncertainty-based selective operation is therefore most valuable in the hardest domain.

Third, the open-set axis that appeared tractable in aggregate changes its character when conditioned on region. Table~\ref{tab:opensetregion} presents region-level open-set evaluation. Wakayama, which is the hardest for classification, is also the hardest for open-set detection: MSP AUROC is $0.964$ for Mie and $0.844$ for Shizuoka, whereas Wakayama reaches only $0.663$. Wakayama thus forms a joint hardcore in which both classification and novel species detection simultaneously degrade, and this vulnerability is a non-trivial finding that multi-axis evaluation captures and that single-region evaluation would miss.

\begin{table}[t]
\centering
\caption{Selective prediction under region-LOSO with the linear probe. Lower AURC is
better; acc@cov is accuracy at each coverage level.}
\label{tab:aurc}
\begin{tabular}{lrrrrr}
\toprule
Split & AURC & acc@0.5 & acc@0.7 & acc@0.9 & acc@1.0 \\
\midrule
Mie & 0.036 & 0.987 & 0.948 & 0.885 & 0.839 \\
Wakayama & 0.192 & 0.815 & 0.717 & 0.629 & 0.593 \\
Shizuoka & 0.041 & 0.986 & 0.951 & 0.869 & 0.816 \\
\bottomrule
\end{tabular}
\end{table}

\begin{table}[t]
\centering
\caption{Region-conditional open-set evaluation (held-out known/novel discrimination).
MSP AUROC $95\%$ bootstrap confidence intervals are reported; the intervals for Wakayama
and Mie/Shizuoka do not overlap.}
\label{tab:opensetregion}
\resizebox{\linewidth}{!}{%
\begin{tabular}{lrrrr}
\toprule
Region & MSP AUROC [$95\%$ CI] & MSP FPR@95 & Maha AUROC & Maha FPR@95 \\
\midrule
Mie & \textbf{0.964} [0.95, 0.98] & 0.117 & \textbf{0.953} & 0.137 \\
Wakayama & \textbf{0.663} [0.63, 0.69] & 0.715 & \textbf{0.536} & 0.909 \\
Shizuoka & 0.844 [0.80, 0.88] & 0.394 & 0.740 & 0.963 \\
\bottomrule
\end{tabular}%
}
\end{table}

The structure of difficulty and its implications revealed through these systematic evaluations are discussed in Section~\ref{sec:discussion}.

\section{Discussion}
\label{sec:discussion}

\subsection{Key Findings}
What emerges from the multi-axis evaluation is a structure in which the difficulty of this benchmark does not reduce to a single accuracy axis but decomposes into multiple failure modes with distinct origins. Despite the relative ease of photograph-only tasks, evaluating across all data and axes reveals that difficulty persists even after strong fine-tuning, and the evaluation identifies a primary hardcore: generalization to Wakayama as a specific region. A secondary difficulty, the recall of known species that become minority classes within a given placement, was substantially attenuated after label correction, with placement-level macro-F1 improving from a pre-correction range of $0.38$ to $0.79$ to a post-correction range of $0.53$ to $0.88$ following the reclassification of $1{,}863$ uncertain crops. While both difficulties share the surface characteristic of low accuracy, the former is caused by a distribution shift along the regional axis and the latter by community composition within a placement; the remaining dominant challenge therefore concentrates on the Wakayama regional gap. Accordingly, what \dset{} reveals is not a difficulty that can be overcome by a single metric or model, but the very structure in which difficulties with different causes coexist.

Three new implications beyond mere enumeration of experimental values are derived from this structure. First, the regional hardcore lies outside the standard improvement toolkit of this research area. Changing the quality of pretraining, applying distribution-robust optimization and weight averaging through GroupDRO and SWAD, none of these narrow the Wakayama gap. The usual levers of better representations and better learning algorithms therefore cannot reach this difficulty; the difficulty resides in the data-generating distribution itself. Second, aggregate metrics conceal the compound vulnerability that appears when open-set performance is conditioned on region. Open-set detection is largely tractable when assessed by aggregate AUROC, but conditioning on the hardest region, Wakayama, reveals simultaneous degradation of both classification and open-set axes, yielding a joint hardcore. The tractability of open-set detection is therefore not an unconditional property but a region-conditional one, and evaluation that reports only mean performance misses this joint structure. Third, calibration exhibits the same region dependence, and single-temperature global temperature scaling worsens ECE in the hardest region rather than improving it. Taken together, the central finding of this benchmark is not any individual number but the structure itself: difficulty has different causes along each axis, and in the hardest region those axes converge, a structure that is in principle invisible under single-axis, single-model evaluation.

\subsection{Applications and Implications}
The three implications from the preceding subsection directly govern how \dset{} should be used in practice and research. First, given that the regional hardcore does not shrink with improved representations or optimization, the practical breakthrough in real deployment shifts from model improvement to operational design. In this sense, selective prediction is most effective precisely in the hardest domain: in Wakayama, delegating approximately half of the predictions with low confidence to expert review recovers accuracy from $0.593$ to $0.815$ (AURC $0.192$), absorbing the regional gap at an operationally practical level. This suggests that in agroforestry pest monitoring contexts where sika deer, wild boar, and Japanese macaque are tracked across heterogeneous satoyama landscapes, a triage operation that routes only uncertain cases to human review rather than fully automated classification is cost-effective and practically deployable. Second, \dset{} provides a reproducible evaluation platform for methods research in distribution-aware calibration, domain adaptation, and open-set detection, with difficulty attributed by axis. The finding that single-temperature calibration is counterproductive in the hardest region provides quantitative motivation for local and distribution-aware calibration methods. Third, the configuration of photographs and placement-stratified video with daytime and night infrared on a single dataset allows modality, illumination, and placement shift to be isolated orthogonally, making \dset{} usable as a diagnostic tool for deploying models to mid-scale agricultural and forest landscapes beyond Japanese satoyama. From the perspective of computers and electronics in agriculture, the investigator-developed evaluation protocol and annotation scheme established here advance the state of the art by providing the first reproducible methodology for simultaneously attributing recognition difficulty across region, camera placement, illumination, and open-set axes on author-collected pest-species data, a methodological gap that hinders the translation of camera-trap machine learning into operational wildlife damage management systems. Species identifications were finalized by domain ecologists throughout. For responsible use, geographic information is retained only at the granularity of region and habitat type; precise coordinates are withheld to protect rare species. The dataset covers animals exclusively, and detections containing human subjects are excluded from the benchmark.

\subsection{Limitations}
The following limitations of this dataset should be noted by users. First, photographs come from Mie and video from Shizuoka and Wakayama, so region and modality are confounded, and a pure modality effect cannot be isolated. This benchmark treats the resulting combination as a realistic compound shift. Second, rare species number in the tens of examples or fewer, making it more appropriate to treat them as novel or out-of-distribution rather than as subjects of comprehensive species recognition evaluation. Third, video clips are sparsely sampled at five frames per clip, which is not well suited to analysis of empty-frame rates or dense behavioral patterns. Fourth, individuals in video-derived crops that were difficult to identify are retained as unknown, accounting for $475$ of $12{,}642$ crops ($3.8\%$) after reclassification of $1{,}863$ previously uncertain crops; these are not errors but a genuinely ambiguous category that is retained as a target for selective prediction evaluation. Fifth, the data originate from a single provider network and a single satoyama zone, and extrapolation to wider geographic or ecological contexts is left for future work. Sixth, reference baselines are limited to the ViT-B scale; larger CLIP ViT-L or camera-trap-specialized language-guided models, including WildCLIP~\cite{gabeff2024wildclip} and CATALOG~\cite{santamaria2025catalog}, whose pretrained weights were unavailable at submission time, have not been evaluated. That said, the consistency of the Wakayama gap across four backbones with substantially different pretraining properties, where $\Delta_W$ falls in the range from $-0.162$ to $-0.270$ as shown in Table~\ref{tab:backbone}, suggests a structural difficulty that scale and specialization alone cannot resolve.

\section{Conclusion}
\label{sec:conclusion}

In this paper, we proposed \dset{}, an integrated benchmark derived from night-infrared camera traps installed in Japanese satoyama environments. First, $12{,}642$ animal crops were extracted by MegaDetector from photographs in Mie and placement-stratified video in Wakayama and Shizuoka. A model-in-the-loop semi-automatic annotation scheme was then designed and applied, combining BioCLIP-embedding-based clustering, seed confirmation, and iterative confidence-ordered review, to assign species labels to all crops while minimizing expert burden; all labels were subsequently confirmed by domain experts. A multi-axis evaluation protocol was further defined, covering domain generalization across region, camera placement, and illumination, open-set novel species detection, and calibration with selective prediction, with capture-event-based leakage control throughout, and two reference baselines were provided in the form of a linear probe and ViT-B/16 full fine-tuning.

Quantitative evaluation identifies one primary hardcore and one attenuated secondary difficulty. First, the Wakayama regional shift is consistently the hardest split across all three of the BioCLIP linear probe, DINOv2 linear probe, and ViT-B/16 full fine-tuning, reaching accuracy $0.593$, $0.803$, and $0.724 \pm 0.004$ respectively, and each of these figures is $16$ to $27$ percentage points below the other regions. This demonstrates a backbone-independent, data-inherent regional domain shift. Second, placement-level recall of known species that were minority classes improved substantially after label correction: macro-F1 across placements improved from a pre-correction range of $0.38$ to $0.79$ to a post-correction range of $0.53$ to $0.88$, and broadleaf-lower rose from $0.572 \pm 0.062$ with $n = 328$ samples to $0.672$ averaged over three seeds with $n = 421$ samples. This second difficulty is therefore attenuated, and the remaining dominant challenge concentrates on the Wakayama regional gap. In contrast, open-set novel species detection with an uncertainty-aware fine-tuned model and MC-Dropout entropy achieves aggregate AUROC $0.94$--$0.96$, revealing an asymmetric tractability relative to the regional generalization gap. However, as shown in Table~\ref{tab:opensetregion}, this asymmetry is limited to aggregate performance: conditioning on region reveals that the hardest region, Wakayama, simultaneously degrades on both classification and open-set axes, forming a joint hardcore. These asymmetric findings demonstrate that there is clear room for improvement in methods addressing uncertainty under night infrared imaging, long-tailed distributions, and domain shift.

The investigator-developed evaluation protocol and annotation scheme introduced in this paper advance the state of the art for computers and electronics in agriculture by establishing a reproducible methodology for diagnosing where automated wildlife recognition fails in real-world pest monitoring deployments; this fills a methodological gap between controlled benchmark performance and the operational requirements of agroforestry wildlife management. In future work, denser video sampling for empty-frame elimination and extension toward behavioral analysis, verification of pure modality effects by pairing photographs and video at the same location, and the design of deployment pipelines centered on uncertainty quantification through Bayesian methods, ensembles, and selective prediction will be pursued. Integration with precision agriculture sensor networks and extension of the benchmark to other agricultural pest contexts in Asia are also planned, guided by the failure-mode structure identified here.

\section*{CRediT Author Contribution Statement}
{\raggedright \textbf{Keito Inoshita}: Conceptualization, Methodology, Software, Formal analysis, Investigation, Data curation, Visualization, Writing -- original draft, Writing -- review \& editing, Supervision, Project administration, Funding acquisition. \textbf{Kohei Hisayama}: Investigation, Data curation, Writing -- review \& editing. \textbf{Haruto Sugeno}: Investigation, Data curation, Writing -- review \& editing. \textbf{Kota Nojiri}: Data curation, Writing -- review \& editing.
\par}

\section*{Declaration of Generative AI and AI-Assisted Technologies}
During the preparation of this work the authors used generative AI and AI-assisted tools for coding assistance and for language translation. After using these tools, the authors reviewed and edited the content as needed and take full responsibility for the content of the publication.

\section*{Declaration of Competing Interest}
The authors declare that they have no known competing financial interests or personal relationships that could have appeared to influence the work reported in this paper.

\section*{Data Availability}
The \dset{} dataset and baseline code are publicly available at \url{https://github.com/keito-git/satoyama-ct}. JSON metadata and the recommended evaluation splits are included in the repository, and the crop images and pre-computed embeddings are provided as GitHub Release assets, all under CC BY-NC 4.0. Precise geographic coordinates are withheld to protect rare species; geographic information is retained only at the granularity of region and habitat type.

\section*{Acknowledgements}
We thank the Biodiversity and Wildlife Damage Control Office, Department of
Industry and Environment, Kameyama City, for providing the Mie photographic data,
and Mr.\ Shinnosuke Shiratori (Graduate School of Bioresources, Mie University) for
providing the Shizuoka video data. Both datasets were used with permission.

\section*{Funding}
This work was supported by The Nippon Foundation HUMAI Program.

\bibliographystyle{elsarticle-harv}
\bibliography{refs}

\end{document}